\documentclass{PoS}
\usepackage{amssymb,amsmath,graphicx}

\let\ifpdf\relax

\title{Neutral Meson Decays into Two Photons from Lattice QCD}
\ShortTitle{Neutral Meson Decays into Two Photons from Lattice QCD}

\author{\speaker{Huey-Wen Lin}\footnote{NT@UW-13-04}\ \ and Saul D. Cohen
\\
Department of Physics, University of Washington, Seattle, WA 98195-1560 \\
}

\abstract{
A precision determination of the neutral-pion width would improve
determinations of the splitting between the up- and down-quark masses, and matrix elements for the decay of neutral mesons into two photons could play a role in the attempt to probe beyond-the-Standard Model physics in muon $g-2$ experiments. The theoretical error is dominated by hadronic light-by-light diagrams, and since direct measurements are extremely difficult, model calculations factorize it into two-photon diagrams connected by the lightest hadrons. We employ perturbative techniques to express the photon as a superposition of QCD eigenstates accessible in lattice-QCD calculations and found that vector-meson dominance is a poor description of the two-photon decay process when both photons are off shell.
}
\FullConference{The X Quark Confinement and the Hadron Spectrum\\
October 7--12, 2012\\
TUM Department of Physics\\
}

\newcommand{\HLbL}{HL$\times$L }

\begin{document}

%%%%%%%%%%%%%%%%%%%%%%%%%%%%%%%%%%%%%%%%%%%%%%%%%%%%%%%%%%%%%%%%%%%%%%%%%%%%%%%
%% MAINMATTER
%%%%%%%%%%%%%%%%%%%%%%%%%%%%%%%%%%%%%%%%%%%%%%%%%%%%%%%%%%%%%%%%%%%%%%%%%%%%%%%

\section{Introduction}

The experimental measurement of the anomalous magnetic moment of the muon came early from CERN and the more-recent BNL E821 tightened the world average to $116592089(63)\times10^{-11}$, providing one of the most accurate measurements of any quantity in particle physics. The future Fermilab experiment E989 intends to improve the precision by another factor of 4 and make the systematic error the same order as the statistical. This could lead to a discovery of the biggest discrepancy from the theoretical Standard Model (SM) value, revealing new physics.
The theoretical Standard Model calculations are complicated, necessitating the inclusion of effects including QED, electroweak and hadronic corrections which give $116591790(65)\times10^{-11}$\cite{Jegerlehner:2009ry}.
This means that there exists a 3--4~$\sigma$ discrepancy between theory and experiment, depending on which theory result is quoted.
Many have supposed that this may be an indication of new physics or that have we miscalculated some part of the Standard Model contributions.

Therefore, we consider all the theoretical contributions order by order in $\alpha$:
the QED and weak contributions are well known, but hadronic vacuum polarization and light-by-light need further work to decrease the remaining theoretical uncertainty.

The QED contribution is well known from Schwinger's one-loop $O(\alpha)$ calculation\cite{Schwinger:1948iu} to modern calculations at 5 loops ($O(\alpha^5)$)\cite{Aoyama:2012wk}.
In the hadronic sector, the leading terms are the hadronic vacuum polarization at $O(\alpha^2)$,
and the hadronic light-by-light contribution, starting from ($\alpha^3$). The hadronic light-by-light (HL$\times$L) gives the biggest uncertainty of all the theoretical contributions, $105(26)\times10^{-11}$.

Since the biggest uncertainties in the SM calculation of $g-2$ are due to the hadronic contributions, there are many on-going efforts to examine them using lattice QCD (LQCD).
These can be grouped into two main directions:
Firstly, many groups have worked on the leading-order hadronic vacuum polarization at $O(\alpha^2)$, including %JLQCD\cite{}, %1206.1375
Aubin et al.\cite{Aubin:2012me}, %PRD86.054509
UKQCD\cite{Boyle:2011hu}, %PRD85.074504
ETMC\cite{Feng:2011ff} %1112.4946
and Mainz\cite{DellaMorte:2012cf} (who have reported their most recent work at this conference). %1203.055

The second direction is to pursue the hadronic light-by-light at $O(\alpha^3)$. One method for addressing \HLbL in LQCD is direct calculation (including QED photon effects explicitly as a dynamical gauge field), as used by RBRC Collaboration (Blum et al.\cite{Blum:2009zz,Blum:2013qu}). However, such a direct calculation of QED on the lattice is very difficult (in part due to strong finite-volume effects), and such techniques have only recently started to see significant signal.
Indirectly, one can study \HLbL by factorization, evaluating the four-point function as an integral over products of two three-point subdiagrams (each containing two photons) connected by a long-lived neutral meson\cite{Knecht:2001qf}. Then, by studying the
neutral-meson--to--two-photons off-shell form factors, we can provide input to model estimations of the \HLbL contribution. We focus on this method in this proceeding.

Many mesonic processes can contribute to the light-by-light factorized diagram, but the pion dominates; in this work, we will address the $\pi^0$ and $\eta_s$ contributions, where $\eta_s$ denotes the purely valence-$s\bar{s}$ part of the $\eta$.
Experimentally, due to the use of transverse-momentum cuts during event reconstruction, only the form factors where of one of the photons is nearly on-shell is measured. This means that in the factorized integral, vector-meson dominance must be assumed to fill out the two-dimensional form factor. We hope to use lattice-QCD calculations to provide inputs for the \HLbL contribution that allow both photons to go off shell. We can use lattice QCD to provide data in missing kinematic regions of the pion form factor and for more mesons than experiment.

Before elaborating on the lattice approach, let us review the experimental situation. The most studied two-photon decay process in experiment is the neutron pion. There are three main classes of experiment: neutral-meson lifetime, photon fusion and Primakoff effect.

The direct neutral-meson lifetime measurement using time of flight of the meson can only measure the form factor with both photons on shell ($Q_1^2=Q_2^2=0$). The most accurate measurement was done at the CERN SPS just waiting for an in-flight pion to decay between foil sheets of varying distances 
\cite{Atherton:1985av}.

Primakoff effect or photoproduction experiments, such as those done by Cornell\cite{Browman:1974cu}, DESY and Tomsk in the 1970s and by PrimEx\cite{Larin:2010kq} in 2010 use the fusion of a nearly real photon emitted as bremsstrahlung from an electron beam off of the electromagnetic field of a heavy nucleus to form a neutral meson. Such experiments are capable of great precision measurement but there have been discrepancies between the measurements of different labs. Although it is possible to measure form factors with one off-shell photon, most experiments merely reported measurements of the width. PrimEx has measured $\pi$, $\eta$, $\eta^\prime$ widths with future plans for form factors.
Primakoff experiments can be precise, but they require nuclear electromagnetic form factors as input to extract the decay width. Given that nuclear EMC effects are still a puzzle, this introduces some uncertainty.

Electron-positron colliders can produce neutral mesons when two photons radiated from the leptons fuse. Such photon-fusion processes have been studied in the CELLO\cite{Behrend:1990sr} and CLEO\cite{Gronberg:1997fj} experiments in the 1990s and more recently by BaBar\cite{Aubert:2009mc,BABAR:2011ad} and Belle\cite{Uehara:2012ag}. Photon fusion experiments can measure off-shell form factors, but only for one off-shell photon. The early experiments at low $Q^2<10\mbox{ GeV}^2$ seemed in good agreement with an asymptotic approach to the perturbative QCD (pQCD) prediction. However, BaBar found that the form factor continued to rise without going to the pQCD value at $Q^2$ as high as 40~GeV$^2$; contrariwise, Belle found results lying between BaBar and pQCD.

%%%%%%%%%%%%%%%%%%%%%%%%%%%%%%%%%%%%%%%%%%%%%%%%%%%%%%%%%%%%%%%%%%%%%%%%%
\section{Two-Photon Decays on the Lattice}

For the meson-to-two-photon decay process, we want to calculate the matrix element
\begin{equation}
\langle \gamma(q_1,\lambda_1)\gamma(q_2,\lambda_2)|\Phi(p) \rangle,
\end{equation}
where $\Phi(p)$ is a neutral-meson operator of momentum $p$ and the two photons have momenta $q_{1,2}$ and polarizations $\lambda_{1,2}$.
An operator carrying the quantum numbers of the photon in lattice QCD (without also including QED fields) will instead create a rho meson (or two pions, depending on the quark mass used). Ji and Jung~\cite{Ji:2001wha} provide an elegant solution for looking at such low-energy matrix elements through lattice-QCD techniques.
Let us start from the continuum path integral of the wanted matrix elements.
We first use perturbative QED to expand in terms of photon fields $A$, coupling to the quark electromagnetic current.
\begin{multline}
\int\!\!{\cal D}A\,{\cal D}\bar{\psi}\,{\cal D}\psi\,e^{iS_{\rm QED}}A^\mu(y)A^\nu(x) \approx \\
\int\!\!{\cal D}A\,{\cal D}\bar{\psi}\,{\cal D}\psi\,e^{iS_0}\left(...+\left[\bar{\psi}\gamma^\rho\psi A_\rho\right](z)\left[\bar{\psi}\gamma^\sigma\psi A_\sigma\right](w)+...\right)A^\mu(y)A^\nu(x)
\end{multline}
Then, we Wick contract the photon fields into propagators. The $\epsilon$'s are the polarization tensors of the photons.
\begin{multline}
-e^2 \lim_{q^\prime \rightarrow q}
\epsilon_\mu^{(1)*} \epsilon_\mu^{(2)*}
			{q_1^\prime}^2 {q_2^\prime}^2 \times \\ \int\!\!d^4x\,d^4y\,d^4w\,d^4z\,
  		e^{iq_1^\prime \cdot x} D^{\mu\rho}(0,z) D^{\nu\sigma}(x,w)
  		\langle 0| T\{j_\rho(z) j_\sigma(w)\} |\Phi(p) \rangle
\end{multline}
Using the explicit form of the photon propagator, most of these integrals go to delta functions, and the above equation simplifies into
\begin{equation}
			e^2 \epsilon_\mu^{(1)*} \epsilon_\mu^{(2)*}
			\int\!\!d^4x\, e^{iq_1 \cdot y}
  		\langle 0| T\{j^\mu(0) j^\nu(y)\} |\Phi(p) \rangle
\end{equation}
Now we can rotate into Euclidean space unless we hit a pole:
\begin{equation}
\frac{e^2 \epsilon_\mu^{(1)} \epsilon_\mu^{(2)}}
         {\frac{Z_\Phi(p)}{2E_\Phi(p)}e^{-E_\Phi(p)(t_f-t)}}
         \int\!\!dt_i\,e^{-\omega_1(t_i-t)} \times
%\\
 \left\langle T\left\{\int\!\!d^3\vec{x}\,e^{-i\vec{p}\cdot\vec{x}}\varphi_\Phi(\vec{x},t_f)
              \int\!\!d^3\vec{y}\,e^{i\vec{q_2}\cdot\vec{y}}j^\nu(\vec{y},t)j^\mu(\vec{0},t_i)  \right\} \right\rangle,
\end{equation}
where the $\{E,Z\}_\Phi(p)$ are the meson energy and operator overlap factor, $\omega_1$ is one of the photon's frequency, which can be chosen arbitrarily (while the the photon's helicity will be fixed through energy conservation), $t_f$ is the Euclidean time when the meson is created.
The nearest pole will be at the $\rho$ mass or a cut will begin at the $\pi\pi$ energy, so we must keep $q^2 < M_\rho^2$ (or $E_{\pi\pi}^2$).

The expression between the angled brackets is just the three-point correlation function with a meson on one end and vector currents at the other end and inserted. This expression we can evaluate on the lattice. The remaining parts describe how to combine QCD states into a photon of the appropriate energy. The most straightforward way to evaluate this is to compute the three-point function on all $t_i$ and perform the integral explicitly.

\begin{figure}[t]
\begin{center}
\includegraphics[width=0.49\textwidth]{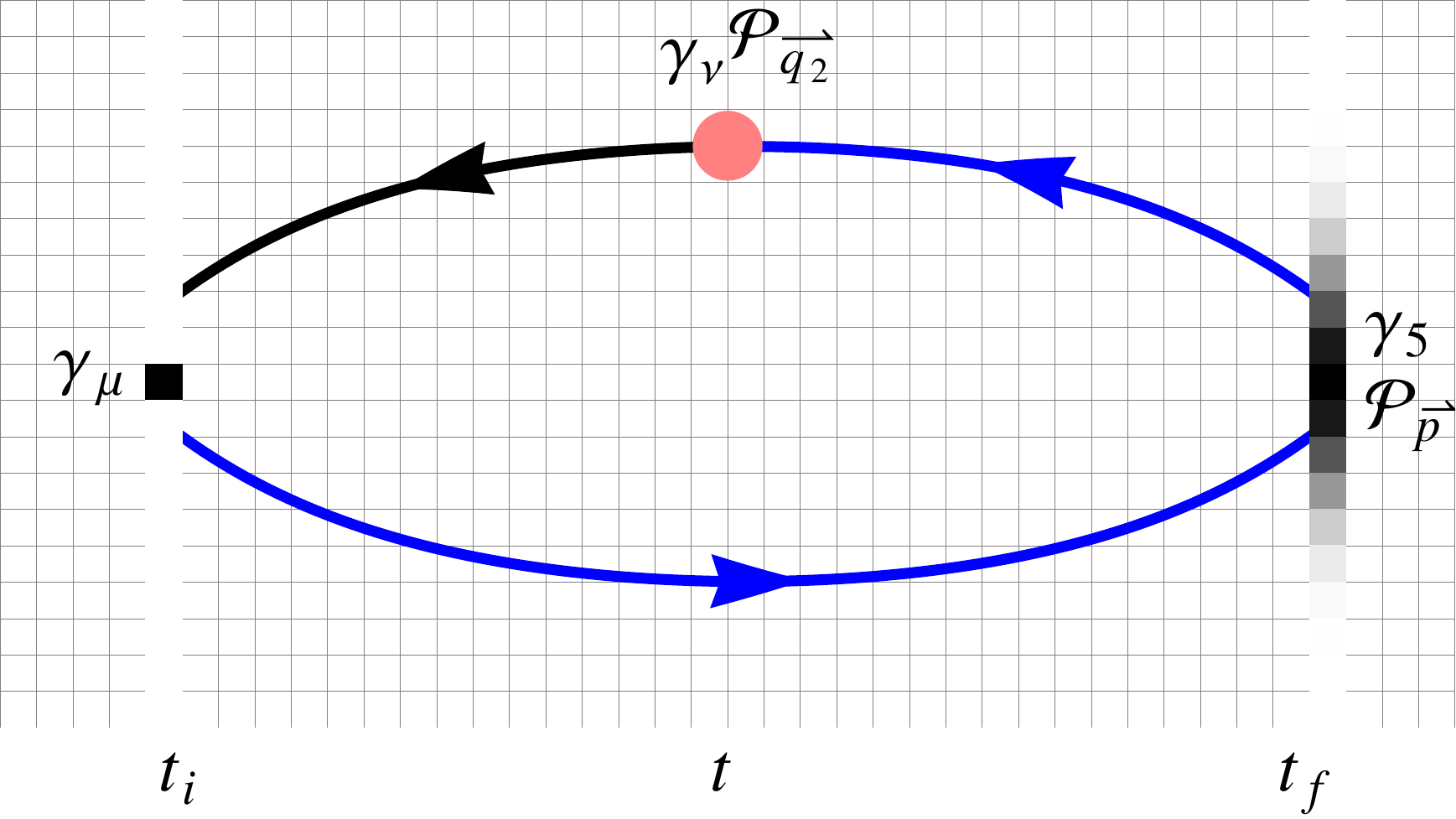}
\includegraphics[width=0.49\textwidth]{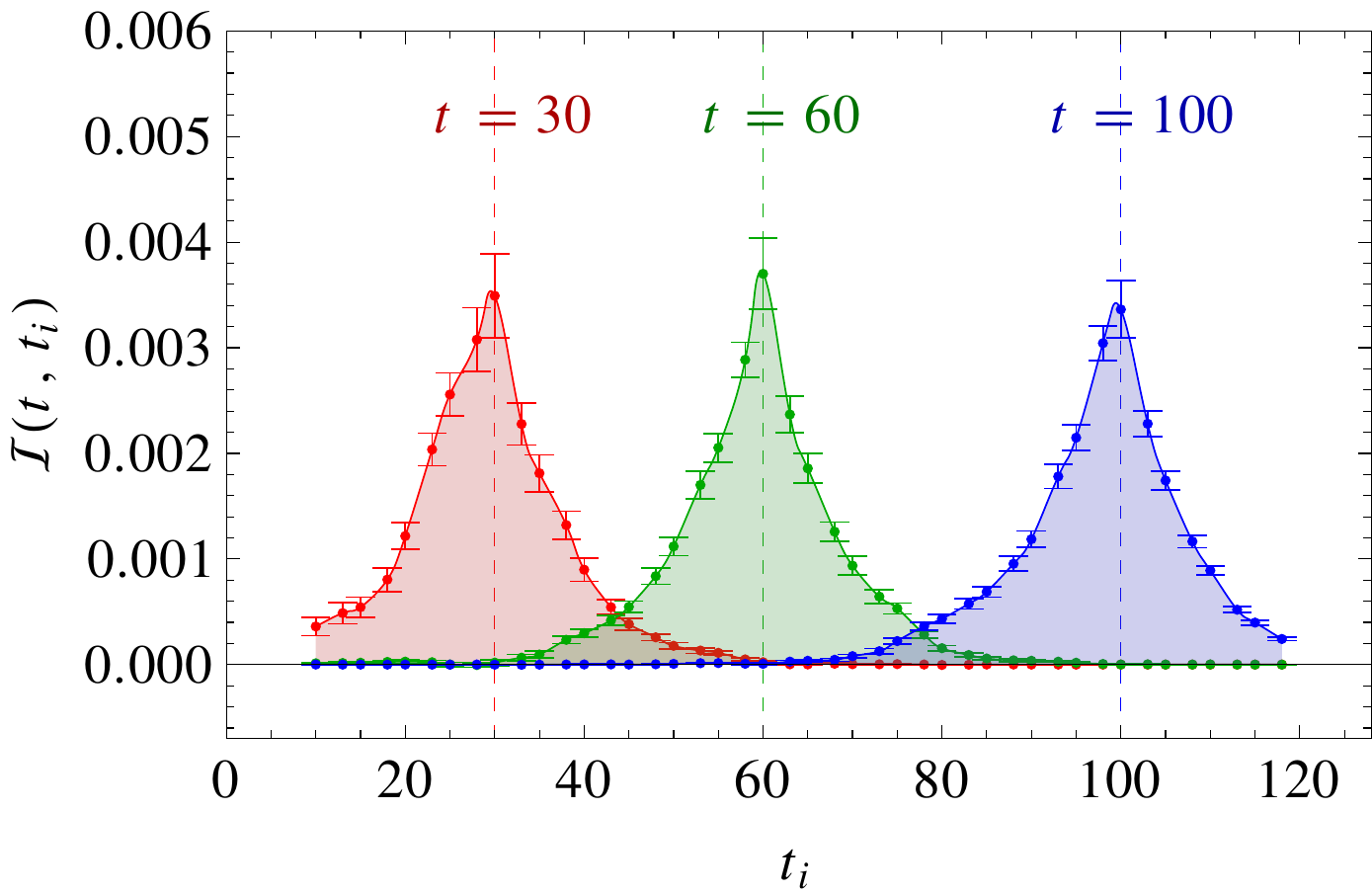}
\end{center}
\caption{\label{fig:1}
(Left) Schematic of the calculation of the lattice three-point correlation function with a smeared meson operator on the right at $t_f$ and point electromagnetic vector-current operators on the left ($t_i$) and center ($t$). Momentum is projected at the meson and current insertion. The first solution of the Dirac equation yields the black quark propagator, while the blue propagator requires a second solution using a sequential source.
(Right) Example of the integrand evaluated from anisotropic lattices at the SU(3) point (with $M_\pi\approx832$~MeV). Three peaks are shown with varying $t$, while the x-axis shows the time to be integrated over, $t_i$. The peaks are free from distortions that may occur due to the Dirichlet boundary and location of the meson operator.
}
\end{figure}

%%% lattice setup

In this work, we concentrate our exploratory work on clover-like actions with the lattice spacing (spatial direction) fixed around 0.12~fm.
We use the Hadron Spectrum Collaboration's anisotropic clover lattice with $M_\pi \in \{830, 560, 450, 390\}$~MeV (100 configurations each)\cite{Lin:2008pr}
and MILC's HISQ lattices with $M_\pi \approx 310$~MeV, $M_{\eta_s} \approx 680$~MeV on 310 and 140~MeV sea pions\cite{MILC:2012uw,Briceno:2012wt}; other lattice spacings are in progress.
For the explicit-integral method, we fix the location of the neutral meson at some timeslice and project onto zero momentum. We generate propagators using the clover action under Dirichlet boundary conditions and apply Gaussian smearing to improve overlap with the ground-state mesons.
We compute a sequential source from a point source at $t_i$, continuing through the smeared source at $t_f$ with momentum projection.
We project onto momenta $0 < |\vec{q_2}|^2 \le 5$ at the insertion at $t$.
The left-hand side of Fig.~\ref{fig:1} shows a demo for the setup.

In order to ascertain whether our calculation will suffer from lattice distortions, we need to scrutinize the time-dependence of the integrand. If the integrand is peaked too sharply, we will not be able integrate it accurately; if it is too wide, we cannot capture the integral within the lattice time extent. In addition, we need to check for distortion due to the boundaries and proximity to the sink timeslice. The peak is well resolved, neither too narrow nor too wide.
Figure~\ref{fig:1} shows an example of the integrand from an anisotropic lattice with $M_\pi \approx $ 850 MeV.
The meson location is fixed at $t_f = 120$ at this example,
and the source is
Gaussian smeared. There is a clearly defined and undistorted peak structure, even at near-boundary times ($t=30, 100$).

\begin{figure}
\begin{center}
\includegraphics[width=0.5\textwidth]{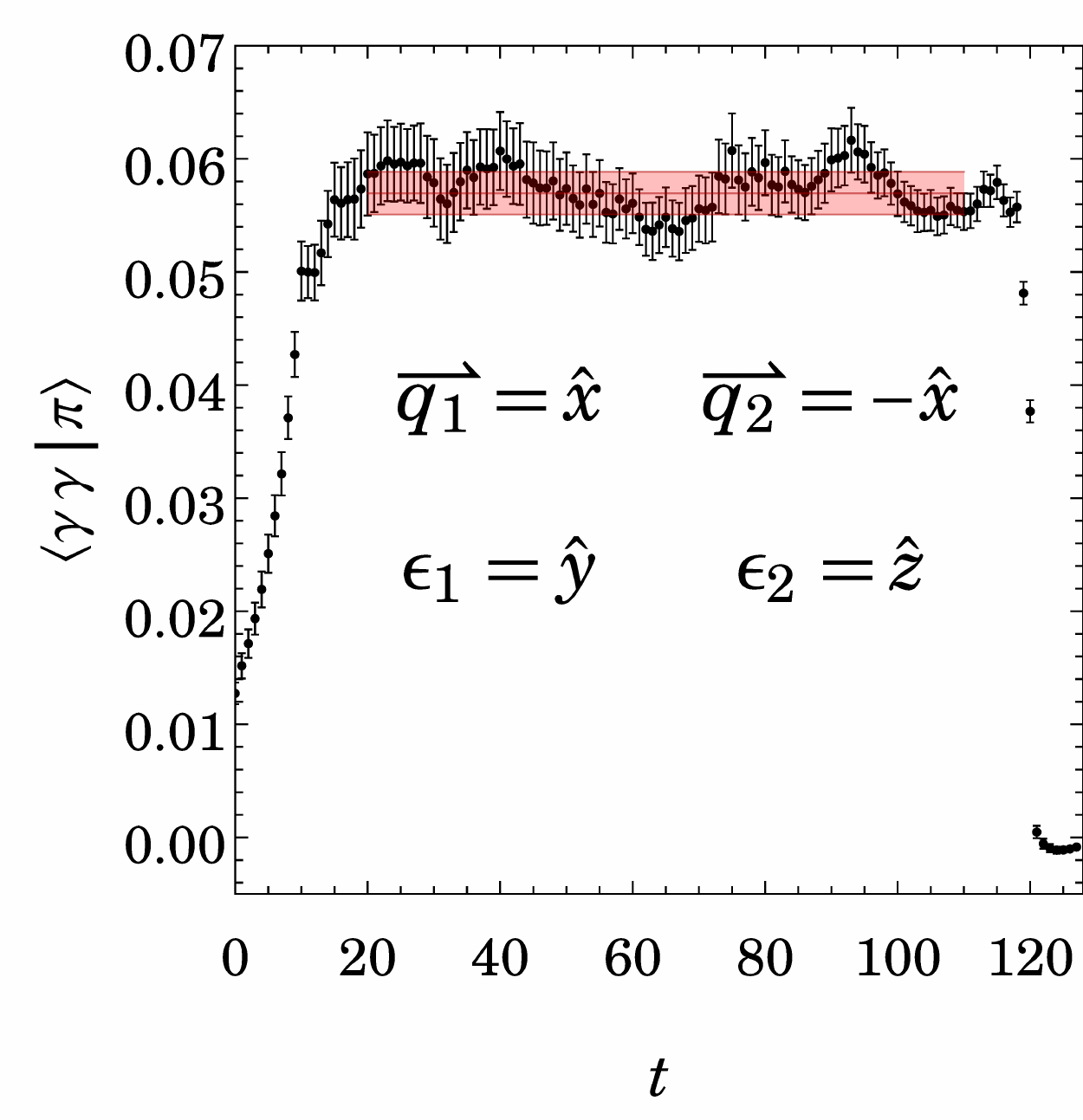}
\includegraphics[width=0.45\textwidth]{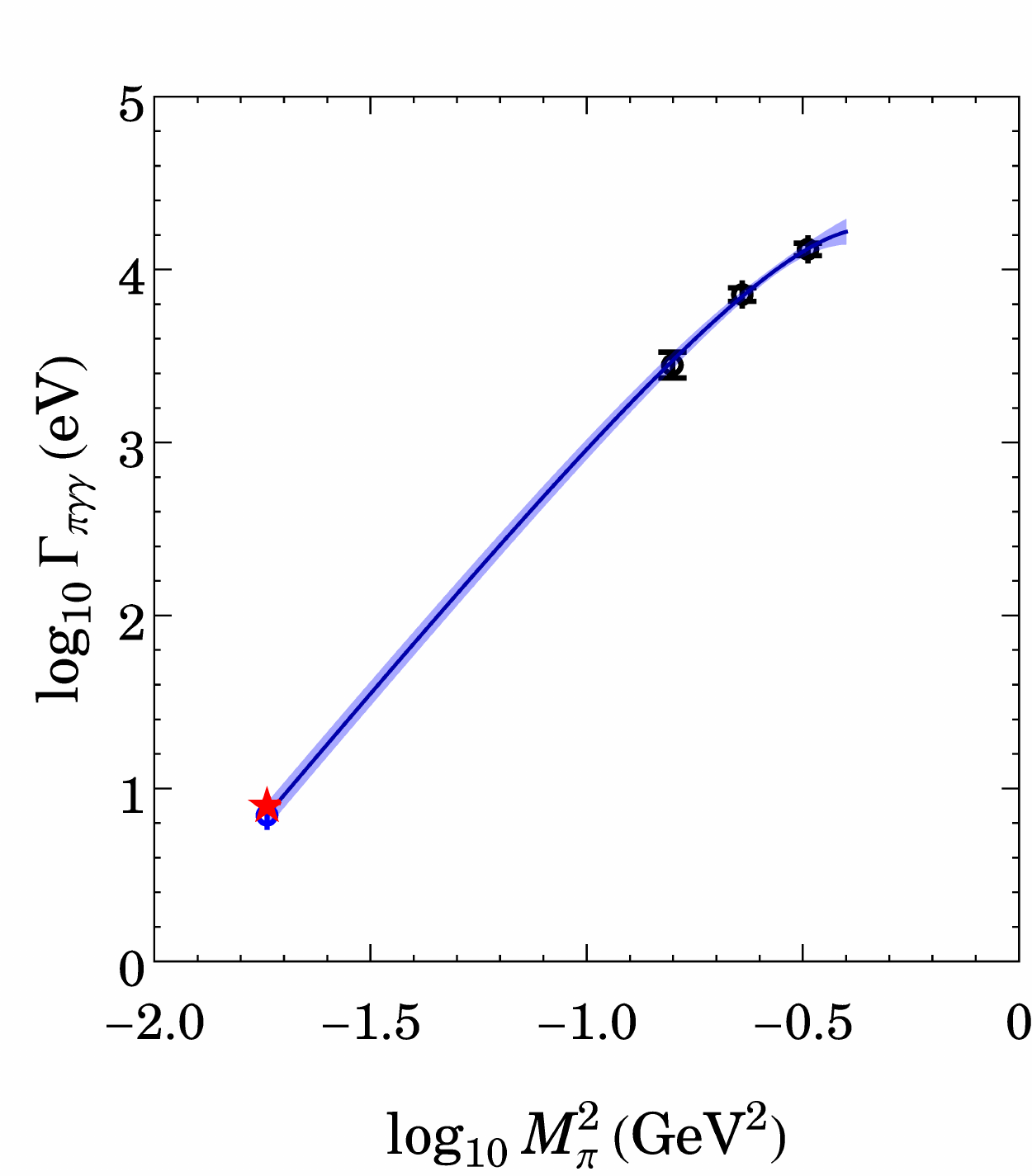}
\end{center}
\caption{\label{fig:2}
(Left) Integrated matrix element for $\vec q_1 \propto\hat{x}$, on the anisotropic lattices at the strange-quark mass. The fitted value over the plateau region is shown as a pink bar.
(Right) Log-log plot of the pion radiative width calculated on anisotropic clover lattices at relatively heavy pion mass ($M_\pi \in\{400,600\}$) (black circles) and a plausible extrapolation using a quartic form (blue), along with the experimental value (red star).
}
\end{figure}

We perform the integral explicitly by summing over all $t_i$ for each value of $t$. Due to kinematic factors, the integral only nonzero when $\varepsilon_{\mu\nu\rho\sigma} \epsilon^\mu \epsilon^\nu q_1^\rho q_2^\sigma \neq 0$. We see a clear plateau in the expected region, away from $t=0$ and $t=t_f$ on the left-hand side of Fig.~\ref{fig:2}, which shows $\vec q_1=\hat x$ for the . Although it is possible for there to be exponential contamination from excited states, no such problem is seen here. This may be due to the large gap between the ground-state pion and its first excited state or due to the excited pion having small coupling to the two-photon state.
All the rotationally equivalent momenta and polarizations at a given set of $Q^2$ are averaged in the final results.

%%%%%%%%%%%%%%%%%%%%%%%%%%%%%%%%%%%%%%%%%%%%%%%%%%%%%%%%%%%%%%%%%%%%%%%%%%%%%%%
\section{Form-Factor Results}

Before we look at the pion radiative form factors, let us start with checking the better known quantity, the pion radiative width $\Gamma_{\pi^0\rightarrow \gamma\gamma}$, at the physical pion mass.
We use the data from anisotropic clover lattices with pion masses below 600~MeV
and a naive extrapolation of the form $aM_\pi^2+bM_\pi^4$. We obtain
 8.7(1.4)~eV (statistical error only) while the best experimental value (from PrimEx) is
 7.82(14)(17)~eV, as shown in the right-hand side of Fig.~\ref{fig:2}.
This may not be a reliable extrapolation given that the pion masses are heavy; however, it gives us some indication that the analysis is on the right track and that the $O(a)$ effects are likely under control.

\begin{figure}
\begin{center}
\includegraphics[width=0.59\textwidth]{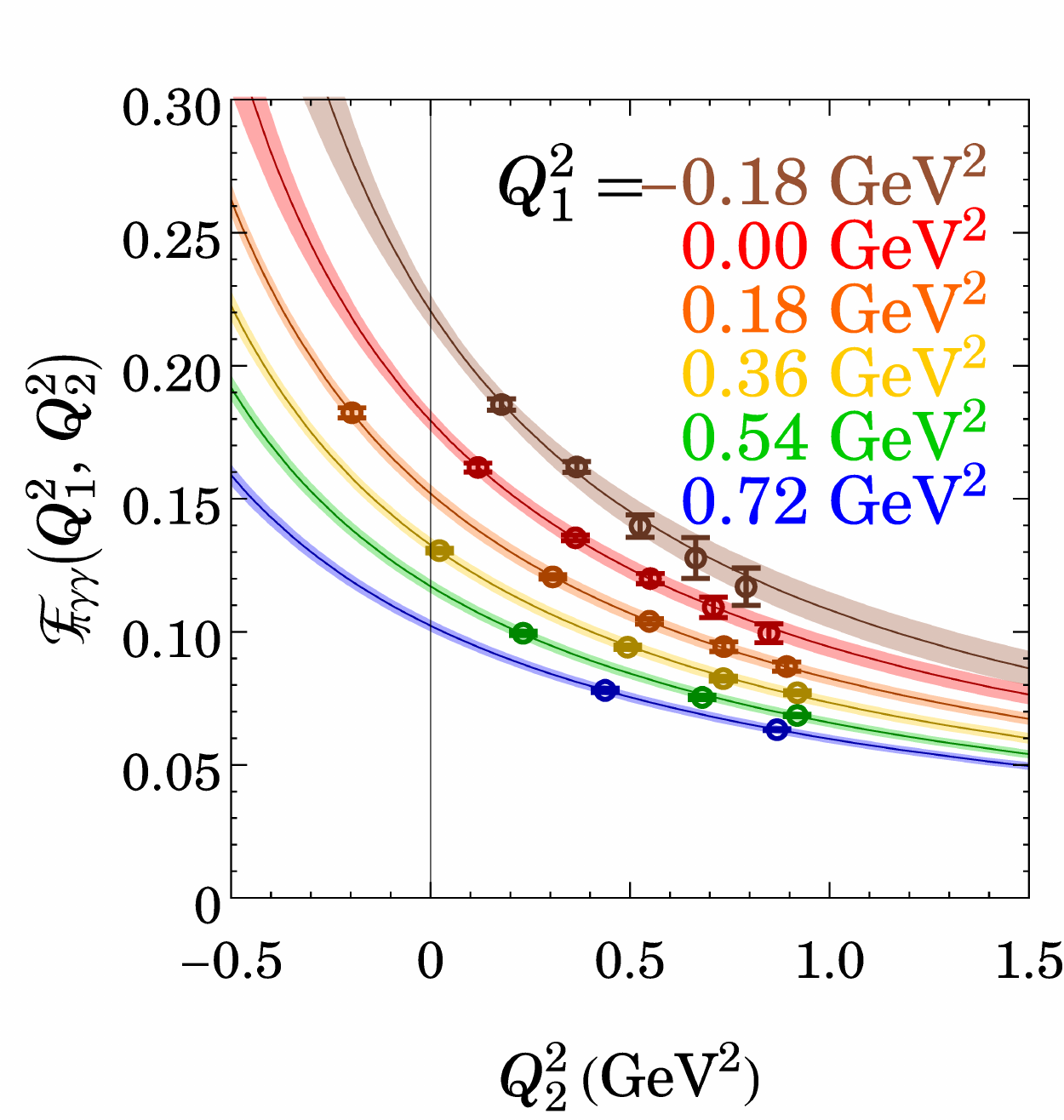}
\includegraphics[width=0.4\textwidth]{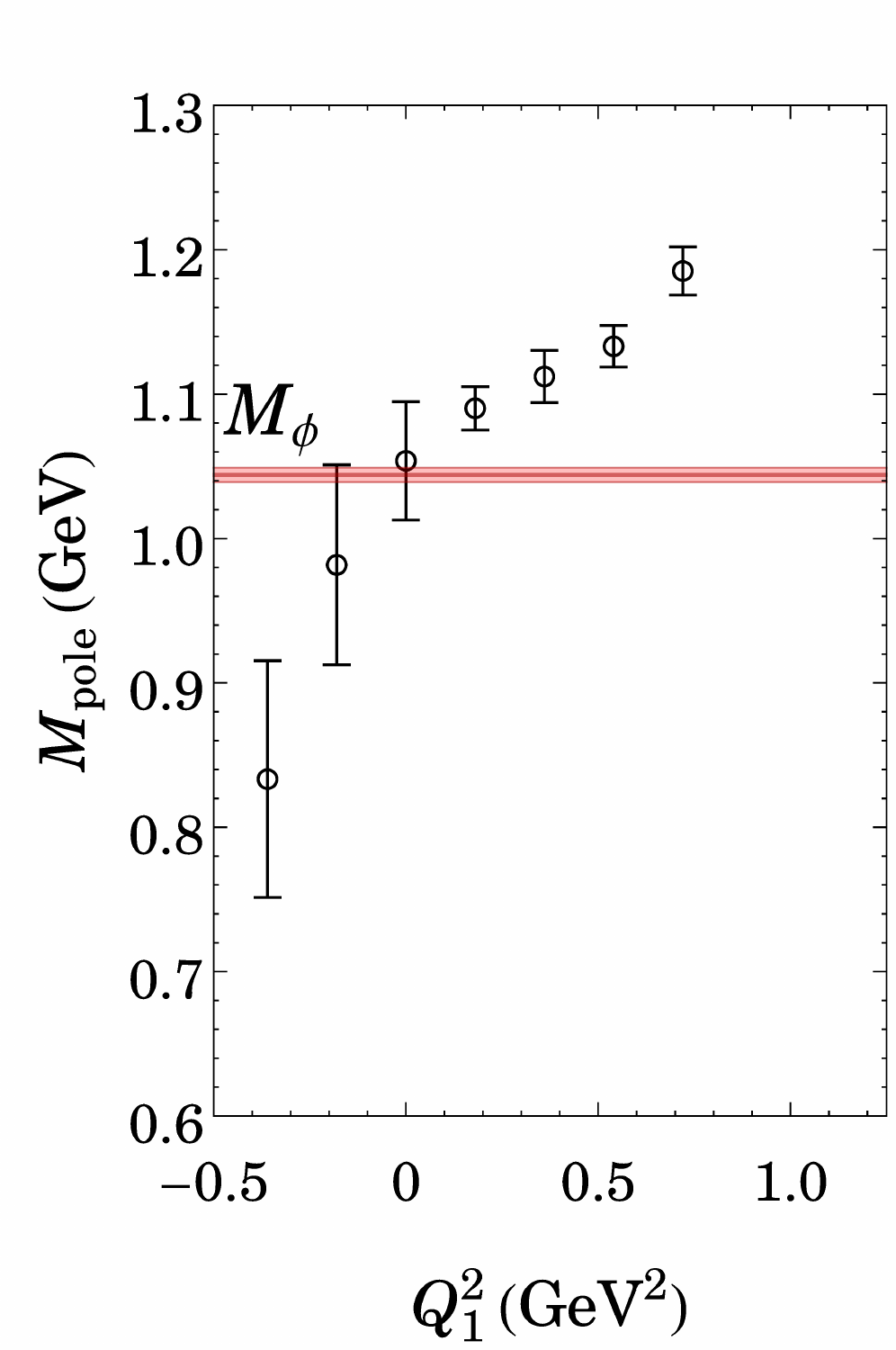}
\end{center}
\caption{\label{fig:3}
(Left) $\eta_s$ form factors from clover on HISQ lattices with pion mass of 310~MeV. The $Q_1^2$ varies from $-0.18$ to 0.72~GeV$^2$. The fits shown are simple monopoles with each $Q_1^2$ being fit separately.
(Right) The extracted pole masses from vector-meson dominance (VMD) extrapolation (black circles) and the vector meson (in this case $\phi$) mass directly calculated (the red line).
}
\end{figure}

Since we are performing the integral explicitly, we can change the weighting factor to set $Q_1^2$ arbitrarily. This means that we are free to explore regions of parameter space where both photons are off shell, which will be useful for comparing to photon fusion experiments and for use in the factorization of light-by-light.
The data are well described by a monopole fit:
\begin{equation}
		{\cal F}(Q_1^2,Q_2^2) = \frac{F(Q_1^2)}{1+Q_2^2/M_{\rm pole}^2(Q_1^2)},
\end{equation}
where we expect from vector-meson dominance that the pole mass should be the vector-meson mass ($\rho$ for light quarks, $\phi$ for strange quarks).
See the left-hand side of Fig.~\ref{fig:3} for an example in the $a=0.12$~fm strange-pseudoscalar ($\eta_s$) meson form factor with $Q_1^2 \in [-0.18,0.72]\mbox{ GeV}^2$ and the monopole fitted bands.

Vector-meson dominance suggests that the pole mass should be approximately equal to the vector-meson mass. We plot $M_{\rm pole}$ as a function of $Q_1^2$ and find that the agreement with the lattice vector-meson mass is relatively good for an on-shell photon but not so good off shell.
The right-hand side of Fig.~\ref{fig:3} shows the fitted pole mass values (black points) for the $\eta_s$ along with the corresponding vector meson $\phi$ measured (the red line) on the same ensemble. 
Note that we obtained similar results for earlier low-statistics data on the CP-PACS clover $a\approx0.09$~fm, $M_\pi\approx725$~MeV~\cite{Cohen:2008ue}, 
as well as other clover on HISQ
$a\approx0.12, 0.09$~fm, $M_\pi\approx310$~MeV. However, we will need higher statistics to see more than one sigma discrepancy.
This suggests that the light-by-light contribution currently being estimated via VMD will require further examination.

In summary, this is an exciting time to explore the limits of the Standard Model with lattice gauge theory. New techniques are allowing us to extend the reach of lattice gauge theory, such as the technique suggested by Ji and Jung, which allows access to electromagnetic quantities with two photons in lattice QCD. 
With sufficient computational resources, lattice calculations can probe photon virtualities inaccessible to experiment. 
Our current results remain limited by low statistics; however, these statistical errors can be reduced by an order of magnitude. 
Preliminary results reproduce the well-measured pion decay width by extrapolating from heavy pion mass ensembles.
Beyond the on-shell form factor, we find that the vector-meson dominance model does not work very well. This implies that more advanced techniques will be needed to apply the factorized integral to the hadronic light-by-light contribution to muon $g-2$.
We intend to further this work by calculating radiative form factors using lighter pion masses and including other neutral mesons, such as the scalar and axial.
Finally, we will investigate the possibility of applying the formalism of Ji and Jung directly to the problem of hadronic light-by-light; this would yield a four-point correlator and multiple integrals, but may yet be easier than the direct QED+QCD method.

%%%%%%%%%%%%%%%%%%%%%%%%%%%%%%%%%%%%%%%%%%%%%%%%%%%%%%%%%%%%%%%%%%%%%%%%%%%%%%%
\section*{Acknowledgments}
HWL would like to thank the conference organizers for the financial support to make the trip.
%%%
The calculations were performed using the Chroma software suite~\cite{Edwards:2004sx}; the parameters were initially tuned on Hyak clusters at the University of Washington managed by the UW Information Technology, using hardware awarded by NSF grant PHY-09227700, and the production was run at FNAL cluster under USQCD SciDAC award. %
The speaker is supported by the DOE grant DE-FG02-97ER4014.

\bibliographystyle{apsrev}
\bibliography{refs}
\end{document}